\documentstyle[11pt]{article}

\makeatletter
\@addtoreset{equation}{section}
\makeatother

\let\a=\alpha \let\b=\beta \let\g=\gamma \let\d=\delta

\let\s=\sigma \let\t=\theta   
      \let\G=\Gamma   
  \let\S=\Sigma   
 
  \let\re=\ref
  
 \def\bd{\begin{document}} \def\ed{\end{document}}
\def\ds{\documentstyle} \let\fr=\frac \let\bl=\bigl \let\br=\bigr
\let\Br=\Bigr \let\Bl=\Bigl
\let\bm=\bibitem

\def\ba{\begin{array}}
\def\ea{\end{array}}

\def\ft#1#2{{\textstyle{{\scriptstyle #1}\over {\scriptstyle #2}}}}
\def\fft#1#2{{#1 \over #2}}
\def\sst#1{{\scriptscriptstyle #1}}
\def\oneone{\rlap 1\mkern4mu{\rm l}}
\newcommand{\ho}[1]{$\, ^{#1}$}
\newcommand{\hoch}[1]{$\, ^{#1}$}
\newcommand{\Str}{\rm Str\, }

\newcommand{\bea}{\begin{eqnarray}}
\newcommand{\eea}{\end{eqnarray}}

\def\[{{[}}
\def\]{{]}}

\newcommand{\be}{\begin{equation}}
\newcommand{\ee}{\end{equation}}

\def\oneone{\rlap 1\mkern4mu{\rm l}}
\newcommand{\p}{\partial}
\newcommand{\tb}{\bar\theta}
\newcommand{\tamphys}{\it Center for Theoretical Physics\\
                      Texas A\&M University, College Station, Texas 77843}
\newcommand{\auth}{\large E. Sezgin\hoch{\dagger}}

\begin{document}

\hfill{CTP TAMU-49/95}

\hfill{hep-th/9512082}

\hfill{December 1995}

\vspace{30pt}

\begin{center}
{ \Large\bf Super p-Form Charges and a Reformulation of the Supermembrane
              Action in Eleven Dimensions\hoch{\ast}}

\vspace{40pt}

\auth

\vspace{15pt}

{\tamphys}

\vspace{50pt}

\underline{ABSTRACT}
\end{center}

We discuss an extension of the super-Poincar\'e algebra in $D=11$
which includes an extra fermionic charge and super two-form charges. We give
a geometrical reformulation of the $D=11$ supermembrane action which is
manifestly invariant under the extended super-Poincar\'e transformations.
Using the same set of transformations, we also reformulate a superstring
action in $D=11$, considered sometime ago by Curtright. While this paper is
primarily a review of a recent work by Bergshoeff and the author, it does
contain some new results.

{\vfill\leftline{}\vfill
\vskip	10pt
\footnoterule
{\footnotesize \hoch{\ast} Talk presented at the Workshop on {\it
      Gauge theories, applied supersymmetry and quantum
gravity}, Leuven, \hfill\break \phantom\quad\quad\quad Belgium, July 10-14,
1995.}\vskip  -12pt}
\vskip 10pt  {\footnotesize\hoch{\dagger}	Research supported in part by
 NSF Grant	PHY-9411543}

\pagebreak
\setcounter{page}{1}

\section{Introduction}

	Developments in the study of nonperturbative string physics over the
last five years have revealed the importance of super $p$--branes. In
particular, there are tantalizing hints at the possibility of an important role
to be played by the eleven dimensional supermembrane theory. There is growing
evidence for the existence of a fundamental theory in eleven dimensions which
is described by the eleven dimensional supergravity in a certain low energy
limit \cite{Witten}. At present, the only available theory of extended objects
that produces correctly the eleven dimensional supergravity seems to be the
eleven dimensional supermembrane theory \cite{BST}. Thus, it is natural to
probe further into the structure of this remarkably unique theory.

	One aspect of the $D=11$ supermembrane theory which is certainly worth
exploring is the structure of its underlying spacetime supersymmetry algebra.
The purpose of this paper is to summarize some recent results obtained by
Bergshoeff and the author \cite{BS/1} in this area, and along the
way, to provide a modest extension of those results. To be more specific, we
will describe the extension of the supersymmetry algebra by the inclusion of an
extra fermionic charge  and super two-form charges. The latter are motivated by
the current algebra that emerges in study of the known supermembrane action.
Using the extended supersymmetry algebra, we shall provide a geometrical
reformulation of the supermembrane action that generalizes a similar result
obtained by Siegel \cite{WS} for the Green-Schwarz superstring. In fact, a
further generalization of our $D=11$ supersymmetry algebra which includes also
super five-form charges seems to be possible. We will give the details of this
extension elsewhere.

The super two-form and five-form charges occurring in the extended
supersymmetry algebra are intimately connected with the existence of
supermembrane \cite{DS} and super five-brane \cite{G} solitons of $D=11$
supergravity, while the role of the extra fermionic generator (which can be
vied as part of a super one-form generator) seems less clear at present. One
might think that it should play a role in the construction of a string theory
in eleven dimensions. Indeed, sometime ago Curtright \cite{TC} constructed a
superstring action in $D=11$. However, the action lacked
$\kappa$--symmetry, and it was not clear if it described a physically viable
theory. Although this still remains to be the case, we will nonetheless
reformulate Curtright's action in a way which makes use of the extended
super-Poincar\'e algebra considered here.

An important motivation for our work is the search for a covariant
supersymmetric action that would describe a super five-brane in eleven
dimensions. A related question is how to describe (preferably in a
covariant fashion) the dynamics of super $p$--brane solitons that arise in
supergravity theories in diverse dimensions \cite{DKL}. Thus, before
presenting our results, we shall first go over various aspects of this problem.

We begin by recalling that super $p$-brane solutions of
supergravity theories seem to be inevitable. For example, in addition to the
membrane and five-brane solitons of $D=11$ supergravity mentioned above, the
heterotic string has string and five-brane solitons, Type IIA supergravity in
$D=10$ has $p$-brane solitons with $p=0,2,4,6,8$ and Type IIB supergravity in
$D=10$ has $p$-brane solitons with $p=-1,1,3,5,7,9$ ($p=-1$ corresponding to
instantons).

	 The super $p$-brane solitons typically preserve half of the
supersymmetries and give rise zero modes that form matter supermultiplets on a
$(p+1)$-dimensional world-volume. One then expects a Nambu-Goto type
supersymmetric actions to describe the dynamics of the zero-mode fields
corresponding to the physical degrees of freedom propagating on the
worldvolume. A natural way to covariantize this action \cite{HLP} is to
introduce worldvolume reparametrization and $\kappa$ symmetries, which then
makes it possible to make the target space Lorentz and supersymmetries
manifest.
However, that is not always a straightforward procedure, if possible at all.

Indeed, the first super $p$-brane action (with $p>1$) was constructed in
\cite{HLP} in a manner described above. A three-brane soliton of $N=2,
D=6$ super-Yang-Mills was considered, and a covariant action describing a super
three-brane in six dimensions was obtained. Later, this result was generalized
to other $p$-branes in various dimensions \cite{BST}, without any reference
to $p$-brane solitons, but solely on the basis of symmetries. While the case of
supermembrane in $D=11$ was emphasized in \cite{BST}, actions and the necessary
constraints on the spacetime background were given for all super $p$-branes
as well. Later, it was shown \cite{AETW} that these constraints restricted the
possible values of $(p,D)$ as follows:  $(p=1; D=3,4,6,10)$, $(p=2;
D=4,5,7,11)$, $(p=3; D=6,8)$, $(p=4; D=9)$ and $(p=5; D=10)$. A common
feature of these theories is that in a physical gauge, the physical degrees of
freedom on the world-volume form scalar supermultiplets.

With the emergence of new kinds of super $p$-brane solitons, it became
clear that there were other possible supermultiplets of zero-modes
on the world-volume. For example, a five-brane soliton in Type IIA
supergravity admits the $(1,1)$ supersymmetric Maxwell multiplet,
and a five-brane soliton in $D=11$ supergravity admits the $(2,0)$
supersymmetric antisymmetric tensor multiplet in $D=6$, as zero-mode
multiplets (see \cite{DKL}, for a review). In fact, one can turn the argument
around and conjecture the existence of super $p$-brane solitons for every
possible matter supermultiplets in $p+1$ dimensions that contains $n$ scalar
fields. Assuming that the sacalars correspond to translational zero-modes, it
follows that the dimension $D$ of the target space is $D=n+p+1$. These leads to
a revised
$p$-brane scan \cite{DL}.

It is useful to make a distinction between super $p$-branes
according to the nature of the word-volume degrees of freedom. We will
tentatively refer to them as $\it scalar$, $\it vector$ and $\it tensor$
{\it $p$-branes}, depending on whether the world-volume physical degrees
of freedom form  scalar, vector or antisymmetric tensor supermultiplets.

The problem of how to describe the dynamics of super $p$--branes of the type
mentioned above seemed rather unsurmountable in the past. However, with
recent  advances in the studies of duality symmetries in nonperturbative
string theory, the prospects of finding a suitable framework for
super $p$--brane dynamics look brighter. In particular, the idea of Dirichlet
$p$--branes \cite{DLP,P} is a very useful one in that it seems to provide a
conformal field theoretic stringy description of the $p$-branes, albeit in a
physical gauge. In this approach, one considers open strings whose boundaries
are constrained to move on $(p+1)$ dimensional plane, which has its own
dynamics, and therefore it can be interpreted as the world-volume of a
$p$-brane. However, this description seems capable of describing only the
scalar
and vector $p$--branes, but not the tensor $p$--branes. Perhaps a
generalization
of the ideas of \cite{P} can also lead to a stringy description of tensor
$p$--branes. This is an open problem.

Another open problem along the lines discussed above is how to find a
{\it covariant} description of the vector and tensor $p$--branes. In this note,
we will summarize some ingredients which might play a useful role
in achieving this. To provide a further background to our work, let us recall
that the original covariant $p$-brane actions of \cite{BST} describe the scalar
$p$--branes and that they are essentially sigma models with Wess-Zumino terms
where the coordinates fields map a bosonic $(p+1)$--dimensional world-volume
$M$
into a target superspace $N$ in $D$--dimensional spacetime. This suggest a
generalization where one considers more general $M$ and $N$. Indeed, models
have been constructed where $M$ itself is elevated into a superspace. For a
formulation of the eleven dimensional supermembrane in such a formalism, see
\cite{PT}. These formulations are rather complicated, however, so much that it
is a nontrivial matter even to show exactly what the physical degrees of
freedom are.

A rather simple reformulation of the superstring \cite{WS} and the
super $p$-branes \cite{BS/1} does exist, however. In this approach, one leaves
the world-volume as bosonic but generalizes the target superspace in such a way
that the target space super Poincar\'e algebra is extended to include new
bosonic and fermionic generators which are motivated by the existence of super
$p$-branes in those dimensions. Such generalizations of the superspace are then
hoped to open new avenues to describe new degrees of freedom. In particular, it
is rather suggestive that the new coordinates introduced in this fashion may
play an important role in the description of duality symmetries in super
$p$-brane theories-- a subject of great potential importance.

With the above motivations in mind, we now proceed to summarize more
explicitly the main ideas involved in the description of the new types of
super $p$-brane actions, based on new types of target space superalgebras.
Although we will focus our attention exclusively on the extension of the
super-Poincar\'e algebra and the supermembrane in $D=11$, the ideas of this
paper can easily be applied to the other super $p$--branes as well
\cite{BS/1}.

\section{ Extension of the $D=11$ Super Poincar\'e Algebra}

The elements of the ordinary super Poincar\'e algebra in $D=11$ are the
translation generator $P^\mu\, (\mu=0,1,...,10)$ and the supersymmetry
generators $Q_\alpha\, (\alpha=1,...,32)$. Let us consider the extra generators
$\S^\a$, $\S^{\mu\nu}=-\S_{\nu\mu}$, $\S^{\mu\a}$ and $\S^{\a\b}=\S^{\b\a}$. We
can show that the following extension of the $D=11$ super Poincar\'e algebra
exists
\bea
\{Q_\a ,Q_\b\} &=& \G_{\a\b}^\mu  P_\mu +
               (\G_{\mu\nu})_{\a\b}\S^{\mu\nu}\ ,\nonumber\\
\[Q_\a, P_\mu \] &=& (\G_\mu)_{\a\b}\S^\b+(\G_{\mu\nu})_{\a\b} \S^{\nu\b}\ ,
                      \nonumber\\
\[ P_\mu, P_\nu\] &=&  (\G_{\mu\nu})_{\a\b} \S^{\a\b}\ , \nonumber\\
\[ P_\mu, \S^{\lambda\tau}\] &=& \ft12
               \delta^{[\lambda}_\mu\G^{\tau]}_{\alpha\beta}\
                 \S^{\a\b}\ , \nonumber\\
\[ Q_\a,\S^{\mu\nu}\] &=& -\ft1{10}
                          \G^{\mu\nu}_{\a\b} \S^\b+(\G^{[\mu})_{\a\b}
                          \S^{\nu]\b}\ ,\nonumber\\
\{Q_\a, \S^{\nu\b}\} &=& \left( \ft14\G^\nu_{\g\d}
      \d^\b_\a + 2 \G^\nu_{\g\a}\d^\b_\d \right) \Sigma^{\g\d}\ ,\label{alg}
\eea
with all the other (anti) commutators vanishing. The generators $\S^{\a\b}$ and
$\S^\a$ are central, and they can be contracted away. In fact, setting $\S^\a$
equal to zero  yields an algebra proposed
recently in \cite{BS/1}, and setting both, $S^\a$ and $\S^{\a\b}$ equal to zero
yields an algebra proposed sometime ago in \cite{BS/3}. A dimensional reduction
of the algebra (\re{alg}) to $D=10$, followed by a chiral truncation and a
suitable contraction, yields an algebra that contains $P_\mu$, $Q_\a$ and
$\S_\a$, found sometime ago by Green \cite{MG}.

The usual Lorentz algebra can be incorporated into (\re{alg}), without spoiling
the Jacobi identities. As the algebra (\re{alg}) is already closed, we shall
not
consider the Lorentz generators any further.

To verify the Jacobi--identities one needs the  following $\G$--matrix
identities
\bea
 && \Gamma^{\mu\nu}_{(\alpha\beta}\Gamma^\nu_{\g\delta)} = 0\ ,
        \nonumber\\
&&\G^\mu_{(\a\b} \G^\mu_{\g\d)} -\ft1{10}\G^{\mu\nu}_{(\a\b}
              \G^{\mu\nu}_{\g\d)}=0\ .   \label{id}
\eea
(Note that, the second identity follows from the first one).

The super two-form $\S$ generators introduced above are motivated by the
structure of a supercurrent algebra that arises in the study of charge density
currents in the known supermembrane action \cite{BS/3}. One expects that
suitable soliton solutions of $D=11$ supergravity to carry the $\S$ charges
\cite{AGIT}. Since, in $D=11$ also a super fivebrane soliton exists
\cite{G}, it is natural to include the fifth rank generators
$\S^{\mu_1...\mu_5}$,  $\S^{\mu_1...\mu_4\a_5}$,..., $\S^{\a_1...\a_5}$. In
fact, we have have already succeeded in including  the $\S^{\mu_1...\mu_5}$ and
$\S^{\mu_1...\mu_4\a_5}$, and a further inclusion of the remaining
$\S$ generators should  be possible. For the purposes of this note, we shall
leave out the super five-form $\S$ generators, which we shall treat elsewhere.

It should be mentioned that extensions of the super-Poincar\'e algebra in
eleven dimensions have been considered before, but they differ from
(\re{alg}). One such extension was given by van Holten and  van Proeyen
\cite{osp}. Their algebra corresponds to $OSp(32|1)$. It contains only 32
fermionic generators $Q_\a$ and all the bosonic generators correspond to
the decomposition of $Sp(32)$ with respect to $SO(10,1)$. A suitable
contraction of this algebra  reduces to an algebra with only $Q_\a$, $P_\mu$
and
$\S_{\mu_1\dots \mu_5}$ kept. In \cite{osp}, also a Lorenzian decomposition of
$OSp(64|1)$ was considered. It contains 64 fermionic generators, and it also
differs from our algebra. In fact, it should be stressed that no connection is
known at present between the algebra (\re{alg}) and any contraction of the
classified simple or semi-simple Lie superalgebras. It would be interesting to
find such a connection.

Another extension of the eleven dimensional algebra was considered by D'Auria
and Fr\'e  \cite{DF}, in their geometrical formulation of the eleven
dimensional
supergravity. Their algebra contains the generators $\S^\a$, $\S^{\mu\nu}$ and
$\S^{\mu_1 \dots \mu_5}$, in addition to $P_\mu$ and $Q_\a$. It can be viewed
as contraction of an extended version of (\re{alg}) that was mentioned above.

Finally, a version of the algebra (\re{alg}) where only $P^\mu, Q^\a$ and
$\S_{\mu_1 \dots \mu_5}$ are kept, has also arisen in \cite{AGIT}, in the
context of a topological extension of the superalgebras for extended objects.
It would be interesting to extend the work of \cite{AGIT} to seek
supermembrane configurations which will carry not only the  bosonic charge
$\S^{\mu\nu}$, but all the charges occurring in (\re{alg}).

Let us now turn to a more detailed discussion of the algebra (\re{alg}). It is
useful to introduce the following notation for the generators
\be
 T_A=\left( P_\mu, Q_\a, \S^\a, \Sigma^{\mu\nu}, \Sigma^{\mu\a},\Sigma^{\a\b}
          \right) , \label{ta}
\ee
and write the superalgebra as $[T_A, T_B\} = f_{AB}{}^C\, T_C$. The structure
constants $f_{AB}{}^C$ can be read off from (\re{alg}). Using these structure
constants, one finds that the Cartan-Killing metric vanishes:
\be
\Str_{\rm adj} (T_A T_B )=0\ .\label{str}
\ee
Presumably, this does not rule out the existence of a suitable nondegenerate
metric, which, however, remains to be constructed.

A suitable parametrization of the supergroup manifold based on the
algebra (\re{alg}) takes the form
\be
U=e^{\phi_{\mu\nu}\Sigma^{\mu\nu}}\ e^{\phi_{\mu\alpha}\Sigma^{\mu\alpha}}
\ e^{\phi_{\a\b}\Sigma^{\a\b}}\ e^{\phi_\a \S^\a}\ e^{x^\mu P_\mu}\
       e^{\theta^\a Q_\a}\ , \label{U}
\ee
where we have introduced the coordinates
\be
     Z^M=\left(x^\mu,
         \theta^\a, \phi_\a, \phi_{\mu\nu},\phi_{\mu\alpha}, \phi_{\a\b}
          \right)\ . \label{Z}
\ee
We can define left--invariant currents $L_i{}^A$ and right--invariant
currents $R_i{}^A$ as usual:
\bea
U^{-1} \partial_i U &=& \partial_i Z^M L_M{}^A T_A
                    =  L_i{}^A T_A\ , \nonumber\\
\partial_i U U^{-1} &=& \partial_i Z^M R_M{}^A T_A
                     = R_i{}^A T_A\ .\label{MC}
\eea
The Maurer--Cartan forms $L^A=dZ^M L_M{}^A$ and $R^A=dZ^M R_M{}^A$ obey the
structure equations $dL^A-\ft12 L^B\wedge L^C\, f_{CB}{}^A=0$
and $dR^A+\ft12 R^B\wedge R^C\, f_{CB}{}^A=0$. (We are using the
conventions of \cite{H}). In particular, we have the Cartan integrable system
\bea
dL^\a &=& 0\ ,\quad\quad dL^\mu = \ft12 L^\a \wedge
         L^\b\G^\mu_{\a\b}\ , \nonumber\\
dL_\a &=& L^\mu\wedge L^\b (\G_\mu)_{\a\b} -\ft1{10} L_{\mu\nu}
          \wedge L^\b \G^{\mu\nu}_{\a\b}\ , \nonumber\\
dL_{\mu\nu} &=& \ft12 L^\a\wedge L^\b (\G_{\mu\nu})_{\a\b}\ ,
                                                          \label{CE}\\
dL_{\mu\a} &=& L^\b\wedge L^\nu (\G_{\mu\nu})_{\b\a} + L^\b\wedge
                L_{\mu\nu}\G^\nu_{\b\a} \ ,\nonumber\\
dL_{\a\b} &=& -\ft12 L^\mu\wedge L^\nu (\G_{\mu\nu})_{\a\b}+\ft12
L_{\mu\nu}\wedge
      L^\mu\G^\nu_{\a\b}  +\ft14 L_{\mu\g}\wedge L^\g \G^\mu_{\a\b}
         + 2 L_{\mu(\a}\wedge L^\g \G^\mu_{\b)\g}\ .\nonumber
\eea

The supergroup generators can be realized as the right--translations on the
group given by $T_A=R_A{}^M\partial_M$, while the supercovariant derivatives
invariant under these transformations can be realized as in terms of
the left--translations as $D_A=L_A{}^M \partial_M$, where $R_A{}^M$ and
$L_A{}^M$ are the inverses of $R_M{}^A$ and $L_M{}^A$, respectively. The
supergroup transformations can be written as
\be
\d Z^M = \epsilon^A R_A{}^M\ , \label{susy/2}
\ee
where $\epsilon^A$ are the constant transformation parameters. The explicit
expressions for $L_i{}^A$ and $\d Z^M$ (with $\S^\a=0$) can be found in
\cite{BS/1}. In particular, including the $S^\a $ generator, one finds that
\footnote{
          In the calculations we never need to raise or lower a spinor index
          using the charge--conjugation matrix. It is convenient to use a
          notation where a given spinor always has an upper or a lower
          spinor--index, e.g.~$Q_\a, \Sigma^{\mu\b}, \theta^\a$, etc. In case
we
          do not denote the spinor indices explicitly, it is always understood
         that they have their standard position, e.g.~$(\G_\mu\t)_\a =
          (\G_\mu)_{\a\b}\t^\b, \tb \G^\mu\p_i\t
          =\t^\a(\G^\mu)_{\a\b}\p_i\theta^\b$, etc.}
\bea
L_i^\a &=& \p_i\theta^\a\ ,\quad\quad  L_i^\mu = \p_i x^\mu +
           \ft12 \tb\G^\mu \p_i \t \ ,\nonumber\\
 L_{i\a} &=& \p_i\phi_\a - \p_i x^\mu \left(\G_{\mu}\t\right)_\a
             +\ft1{10}\p_i \phi_{\mu\nu}\left(\G^{\mu\nu}\t\right)_\a
            -\ft16\left(\G_{\mu}\t\right)_\a\tb\G^\mu\p_i\t\ ,\nonumber\\
L_{i\mu\nu} &=& \p_i \phi_{\mu\nu} +\ft12\tb\G_{\mu\nu}
                   \p_i \t\ ,\nonumber\\
L_{i\mu\a} &=& \p_i\phi_{\mu\a}+\p_i \phi_{\mu\nu}\left(\G^\nu\t\right)_\a
               +\p_i x^\nu \left(\G_{\mu\nu}\t\right)_\a +
               \ft16\left(\G_{\mu\nu}\t\right)_\a \tb
               \G^\nu\p_i\t +\ft16\left(\G^\nu\t\right)_\a \tb
               \G_{\mu\nu}\p_i\t\ , \nonumber\\
L_{i\a\b} &=& \p_i \phi_{\a\b}-\ft12 x^\mu\p_i \phi_{\mu\nu}(\G^\nu)_{\a\b}
              +\p_i \phi_{\mu\nu}\left(\G^\mu\t\right)_{(\a}
              \left(\G^\nu\theta\right)_{\b)} +\ft14
              \left( \tb \p_i \phi_\mu\right) (\G^\mu)_{\a\b}\nonumber\\
   && +2\left(\G^\mu\t\right)_{(\a}\p_i\phi_{\mu\b)}-\ft12 x^\mu\p_i x^\nu
      (\G_{\mu\nu})_{\a\b} -\left(\G^\nu\t\right)_{(\a} \left(\G_{\mu\nu}\t
       \right)_{\b)}\p_i x^\mu \nonumber\\
   && -\ft1{12}\left(\G_\nu\t\right)_{(\a}\left(\G^{\mu\nu}\t
       \right)_{\b)}\left(\tb\G_\mu\p_i\t\right)
      -\ft1{12}\left(\G_\nu\t\right)_{(\a} \left( \G_\mu\t
        \right)_{\b)}\left(\tb\G^{\mu\nu}\p_i\t\right)\ .\label{L}
\eea

Just as in the string case \cite{WS}, since the $\S$ generators (anti) commute
with each other, one can consistently impose the physical state condition
$\S\ \Phi (Z)=0$, leading to superfields $\Phi(x,\theta)$ that depend only on
the coordinates of the ordinary superspace $(x^\mu,\theta^\a)$. Of course,
there may be subtleties in imposing these conditions which may arise from
global considerations which may lead to  drastically  different
quantization schemes. We leave these questions for a future investigation.

\section{ Reformulation of the $D=11$ Supermembrane Action}

The usual formulation of the $D=11$ supermembrane action, as well as its
reformulation (in flat as well as curved superspace) takes the following
universal form \cite{BST}
\bea
  I=\int d^3 \s && \bigg[ -\ft12 \sqrt {-\gamma}
           \gamma^{ij}\left(\partial_i Z^M\, L_M{}^a \right)
	         \left(\partial_i Z^M\, L_M{}_a \right)+\ft12\sqrt {-\gamma}
\nonumber\\
           &&-\epsilon^{ijk} \partial_i Z^M \partial_j Z^N \partial_k Z^P\,
             B_{PNM}\bigg] \ , \label{I}
\eea
where $a=0,1,...,10$ is the tangent space Lorentz vector index, $\gamma_{ij}$
is the worldvolume metric and $\gamma={\rm det}\gamma_{ij}$. The superspace
coordinates $Z^M$ and the super three-form $B_{MNP}$ have to be defined in each
formulation. In the usual formulation, the superspace coordinates are
$Z^M=(x^\mu,\t^\a)$, corresponding to the ordinary $D=11$ super-Poincar\'e
algebra, and $B$ is defined in such a way that its field strength $H=dB$
satisfies certain superspace constraints which can be found in \cite{BST}.
These constraints, along with super torsion constraints ensure the
$\kappa$--symmetry of the action. In flat superspace, $H$ takes the form
\cite{BST}
\be
  H= L^\mu \wedge L^\nu \wedge L^\a \wedge L^\b\,
          \G_{\mu\nu\a\b}\ .\label{H}
\ee
One can show that $dH=0$, by using the first two equations in
(\re{CE}) and the $\G$--matrix identity (\re{id}). Furthermore, one can
solve for $B$ as follows \cite{BST/2}:
\bea
B_{\rm old} = && L^\mu \wedge L^\nu\wedge L^\a
            (\G_{\mu\nu}\t)_\a -\ft12 L^\mu\wedge L^\a\wedge L^\b
            (\G_{\mu\nu}\t)_\a (\G^\nu\t)_\b \nonumber\\
&&  -\ft1{12} L^\a\wedge L^\b\wedge L^\g (\G_{\mu\nu}\t)_\a
           (\G^\mu\t)_\b (\G^\nu)\t)_\g \ . \label{old}
\eea

In the new formulation of the supermembrane \cite{BS/1}, the coordinates $Z^M$
now refer to those defined in (\re{Z}). In order to maintain the
$\kappa$ symmetry, the field strength $H=dB$ should still
take the form (\re{H}). However, the Bianchi identity $dH=0$ must now be
satisfied in the full supergroup whose algebra is given in (\re{alg}). Indeed,
using the structure equations of the full group as given in (\re{CE}), we have
shown that $dH=0$, and that the corresponding super three-form takes
the following $G_L$ invariant form \cite{BS/1}:
\be
 B_{\rm new} =\ft23 L^\mu\wedge L^\nu\wedge L_{\mu\nu} +\ft35
              L^\mu\wedge L^\a \wedge L_{\mu\a} -\ft2{15} L^\a \wedge
              L^\b\wedge L_{\a\b}\ . \label{B/2}
\ee
The fact that $H$ takes the same form in both formulation means that the
dependence on all of the $\phi$ coordinates associated with the $\S$
charges is contained in total derivative terms. An interesting fact is that,
the central generators $\S^{\a\b}$ are essential for  this phenomenon to
happen.
Indeed, one can show that the last term in (\re{B/2}) is necessary for $H=dB$
to
take the required form (\re{H}). All these results have stringy analogs, as
discovered by Siegel \cite{WS}.

The action (\re{I}) is manifestly invariant under the global $G_L$
transformations, which include the supersymmetry transformations \cite{BS/1}.
There is a Noether current associated with this global symmetry. The algebra of
the corresponding charge densities contains field dependent extensions, as
expected. The action is also invariant under the local $\kappa$--symmetry
transformations
\be
\d Z^M=\kappa^\a (1+\G)_\a{}^\b L_\b{}^M\ ,  \label{kappa}
\ee
where $Z^M$ are the full superspace coordinates (see (\re{Z})), and $\G$ is
defined by
\be
\G ={1\over 3!\sqrt{ -\g}} \epsilon^{ijk} L_i^\mu L_j^\nu L_k^\rho
\Gamma_{\mu\nu\rho}\ .\label{G}
\ee

One might consider the possibility of using a closed four-form that
would differ from (\re{H}) by containing the left-invariant one-forms other
than $L^\mu$ and $L^\a$. Indeed, we have found two simple such forms, which we
denote by $H'$ and $H''$. They are given by
\bea
H' &=& L^\a \wedge L^\b \wedge L_{\mu\nu} \wedge L^\nu\, \G^\mu_{\a\b}\ ,
   \label{H1}\\
H'' &=& L^\a \wedge L^\b \wedge L_{\mu\g} \wedge L^\g\, \G^\mu_{\a\b}\ .
   \label{H2}
\eea
Writing $H'=dB'$ and $H''=dB''$, one finds that
\bea
 B' &=& -\ft23 L^\mu\wedge L^\nu\wedge L_{\mu\nu} + \ft3{10}
              L^\mu\wedge L^\a \wedge L_{\mu\a} -\ft1{15} L^\a \wedge
              L^\b\wedge L_{\a\b}\ , \label{BN1}\\
B'' &=&  \ft15 L^\mu\wedge L^\a \wedge L_{\mu\a} + \ft25 L^\a \wedge
              L^\b\wedge L_{\a\b}\ . \label{BN2}
\eea
We can use a combination of $B_{\rm new}$, $B'$ and $B''$ to construct a new
supermembrane action. If the combination used contains $L_{\mu\nu}$ and/or
$L_{\mu\a}$, then we need to introduce kinetic terms for the new coordinates
$\phi_{\mu\nu}$ and/or $\phi_{\mu\a}$. It is a nontrivial matter to achieve
$\kappa$ symmetry, or its analogs, in such actions. Ideally, one would like
to find such symmetries to gauge away the unwanted degrees of freedom, and to
arrive at an anomaly-free consistent theory without ghosts and tachyons.

Interestingly enough, Curtright \cite{TC} did consider a superstring
theory in eleven dimensions which contained the extra coordinates
$\phi_{\mu\nu}$. Although, it is not clear how to achieve in this model the
properties mentioned above, it is nonetheless interesting to see how it can be
reformulated in our geometrical framework, based on the  extended
super-Poincar\'e algebra that contains both, the extra fermionic charge and
the super two-form charges. We now turn to a description of this model.

\section{Superstring in Eleven Dimensions?}

Let us assume that a Green-Shwarz type action for superstring in eleven
dimensions consist of a kinetic term and a Wess-Zumino term. The latter would
require the existence of a closed super three-form  in target superspace. Given
the ingredients of the geometrical framework described in the previuos
sections, we see that indeed such a form exists:
\be
    H_3=L^\a \wedge L^\b \wedge \left(  L^\mu (\G_\mu)_{\a\b} -\ft1{10}
                L_{\mu\nu}\G^{\mu\nu}_{\a\b}\right) \ . \label{H3}
\ee
This form is closed, due to the $\G$--matrix identities (\re{id}). In fact, it
is easy to see that we can write $H_3=dB_2$, with
\be
    B_2=L_\a\wedge L^\a\ .  \label{B3}
\ee
Since $H_3$ depends on $L_{\mu\nu}$, we need to introduce a kinetic term for
the coordinates $\phi_{\mu\nu}$. The simplest choice for a manifestly
supersymmetric action is then
\be
 I=\int d^2 \s \bigg[ -\ft12 \sqrt {-\gamma}
           \gamma^{ij}\left(L_i^\mu L_{j\mu}+L_i^{\mu\nu}L_{j\mu\nu}\right)
	         -\epsilon^{ij} \partial_i Z^M
          \partial_j Z^N \,  B_{NM}\bigg] \ , \label{I2}
\ee
where $Z^M=(x^\mu, \t^\a, \phi_\a, \phi_{\mu\nu})$,
$L_i^\mu$ and $L_i^{\mu\nu}$ are defined
in (\re{L}) and $B_{NM}$ are the components of the super two-form $B_2$ defined
in (\re{B3}).  Dropping the total derivative term that contains the
coordinate $\phi_\a$, the action (\re{I2}) reduces to
\be
I=\int d^2 \s \bigg[ -\ft12 \sqrt {-\gamma}
           \gamma^{ij}\left(L_i^\mu L_{j\mu}+L_i^{\mu\nu}L_{j\mu\nu}\right)
	         +\epsilon^{ij}\tb\big( L_i^\mu\G_\mu
    -\ft1{10} L_i^{\mu\nu}\G_{\mu\nu}\big)\p_j\theta \bigg] \ . \label{I2C}
\ee
The Nambu-Goto version of this action where the kinetic terms are replaced by
$\sqrt {-\g}$ with $\g_{ij}=L_i^\mu L_{j\mu}-\ft1{10}
L_i^{\mu\nu}L_{j\mu\nu}$ was proposed long ago by Curtright \cite{TC}.
Considering local fermionic transformations of the form $\d Z^M=\kappa^\a
L_\a{}^M$, one finds that invariance of the action under these
transformations imposes the condition, $P_-^{ij}(L_j^\mu\G_\mu -\ft1{10}
L_j^{\mu\nu}\G_{\mu\nu})\kappa=0$, where $P_-^{ij}=(\g^{ij}-\epsilon^{ij}/\sqrt
{-\g})$. This is a very stringent condition on the parameter $\kappa$, and we
can find no solution in eleven dimensions. Furthemore, as
discussed in \cite{TC}, the physical significance of this action is not clear.
It is conceivable that new kinds of fermionic and bosonic local symmetries that
generalize the $\kappa$--symmetry exist in the enlarged superspace and that
they are crucial in determining the true degrees of freedom and in finding a
physically viable model. To find such symmetries, a better geometrical
understanding of the $\kappa$--symmetry is needed.

\section*{Acknowledgements}

I thank E. Bergshoeff, T. Curtright, M.J. Duff,
M. G\"unayd\i n and W. Siegel for discussions.

\pagebreak

\end{document}